# Unlocking klockmannite: formation of colloidal quasi-2D CuSe nanocrystals and photo-physical properties arising from crystal anisotropy

Urvi Parekh[1], Nadiia Didukh[1], Samira Dabelstein[1], Ronja Piehler[1], Eugen Klein[1], Jivesh Kaushal[1], Tobias Korn[1], Stefan Lochbrunner[1,2], Christian Klinke[1,2], Stefan Scheel[1]*, Rostyslav Lesyuk[1,3]*

[1] Institute of Physics, University of Rostock, Albert-Einstein-Straße 23, 18059 Rostock, Germany
[2] Department "Life, Light & Matter", University of Rostock, Albert-Einstein-Straße 25, 18059 Rostock, Germany
[3] Pidstryhach Institute for applied problems of mechanics and mathematics of NAS of Ukraine, Naukowa str. 3b, 79060 Lviv, Ukraine

**Abstract**

Copper selenide is an exceptional quasi-layered monolithic material that exhibits both semiconducting and metallic properties in adjacent visible and near-infrared (NIR) spectral ranges. Here we introduce a thiol-free colloidal synthesis for generating quasi-2D klockmannite copper selenide nanocrystals via hot injection method, achieving shape control by tuning the injection temperature and precursor concentrations without any additional ligands. This approach produces large klockmannite nanosheets with lateral sizes from 200 nm to several micrometres, as well as uniform triangular nanoplatelets with sizes of 12–25 nm that are monocrystalline and display strong NIR plasmonic absorption. The spectral features of the anisotropic klockmannite phase in the NIR have been analysed using complex-scaled discrete dipole approximation (CSDDA) calculations, which reveal pronounced optical anisotropy and the emergence of hyperbolic regime. The combined effect of propagating and evanescent fields is regarded as the underlying reason of such modes in the hyperbolic domain. Finally, the ultrafast photophysical behaviour of the material in klockmannite phase is examined, including hot-hole cooling, trapping, and coherent phonons generation. Our findings emphasize the important role of the intrinsic crystal anisotropy in governing the physical properties of nanoscale klockmannite.

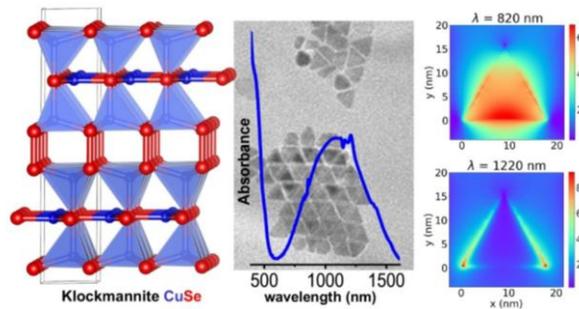

---

* corresponding authors: stefan.scheel@uni-rostock.de, rostyslav.lesyuk@uni-rostock.de

## Introduction

In recent years, metal chalcogenides have garnered strong attention as inorganic semiconductors due to their potential use in solar cells, photodetectors, thermoelectric converters, photocatalysts, flexible electronics and a wide range of other optoelectronic applications [1–7]. The main constituents of groups II-VI, IV-VI, and III-V metal chalcogenides often contain toxic or rare elements. Materials including $Cu_2S$, $CuInS_2$, $CuInSe_2$, CIGS and CZTS have long been used as thin-film absorbers[8,9], and their colloidal nanocrystals[10] (NCs) have more recently enabled solution-processable devices, conductive films, and photothermal or photocatalytic applications in various media[11–17]. Also, copper selenides ($Cu_xSe$) are among such significant metal chalcogenides, comprising a variety of stoichiometric and non-stoichiometric phases, including cubic ($Cu_2Se$ or $Cu_{2-x}Se$), monoclinic ($Cu_2Se$), tetragonal ($Cu_3Se_2$), hexagonal klockmannite (CuSe or $Cu_{0.87}Se$), orthorhombic ($Cu_5Se_4$ or CuSe) and marcasite ($CuSe_2$)[18] phases. All these phases exhibit p-type semiconductor behavior due to the presence of copper vacancies in their crystal lattice with bandgaps covering the range of 1.2 to 2.4 eV[19].

These copper selenides also exhibit tunable localized surface plasmon resonances (LSPR) arising from high hole concentrations, with the strongest LSPR in $Cu_{2-x}Se$ ($0<x\leq1$) non-stoichiometric compositions[20]. LSPRs for these materials are observed in the near-infrared (NIR) region similar to copper sulphides [21], enabling their applications in various fields such as fiber optical communication in the NIR transparency window or as bio-compatible materials. Similar to metal nanoparticles, the observed LSPRs of degenerately doped semiconductor NCs (for instance, $Cu_xSe$) can be tuned via particle size[22], shape[23], and surface chemistry as well as by modulating the free carrier density and distribution within the NPs[24,25]. Since the plasmonic characteristics of these nanomaterials are dependent on their size and shape, which in turn are influenced by the experimental procedure used, the systematic optimization of synthetic conditions is essential to obtain the desirable plasmonic properties.

The compositions and crystal structures of copper selenide NCs are crucial determinants of their properties. Various chemical approaches have been explored to synthesise 2D $Cu_xSe$ NCs. Notably, many reported routes rely on selenium precursors requiring elevated decomposition temperatures, such as trioctylphosphine-Se (TOP-Se)[26,27], which favour thermodynamically stable phases[28]. However, in comparison with $Cu_2Se$ and $Cu_{2-x}Se$, relatively few studies have focused on 2D CuSe, despite its distinct crystal structure and highest concentration of quasi-free holes among copper selenides. This distinction is important because $Cu_{2-x}Se$ has a non-layered structure with vacancy-driven plasmonics, whereas CuSe possesses a quasi-*layered* klockmannite structure that provides intrinsic hole density, strong dielectric anisotropy, and the potential for hyperbolic optical behaviour not accessible in $Cu_{2-x}Se$. Several solution syntheses have been reported for quasi-2D CuSe. Among them, OLA-only syntheses are particularly notable[29,30], as oleylamine can act as both solvent and reductant; however, these routes frequently yield mixed $Cu_{2-x}Se$/CuSe phases or require additional $Cu^+$ injection to preserve nanosheet morphology, limiting phase purity and shape control[31]. Other strategies include ionic-liquid Se precursors[32], microwave-assisted syntheses producing broad lateral distributions[33,34], and PVP-assisted hot-injection methods[4].

The synthesis of single-phase CuSe is inherently challenging due to its complicated klockmannite structure and the intrinsic variable valence states of Cu and Se, compared to the $Cu_{2-x}Se$ solution synthesis[35]. The klockmannite structure features alternating covalently bonded $CuSe_3$-$Cu_3Se$-$CuSe_3$ layers with hypothesized Se–Se van der Waals layers along the z-axis[31] or covalent bonds between chalcogens in 4e sites.[36] Thus, the bond valence of Cu with Se at all lattice sites remains debated. This structural complexity makes CuSe substantially more difficult to stabilise than $Cu_{2-x}Se$, which forms readily under similar conditions. To the best of our knowledge, a synthesis approach that efficiently produces CuSe NCs with uniform shape or size control is still lacking. Thus, an eco-friendly, facile and effective solution method for synthesizing high-quality klockmannite CuSe nanostructures is needed.

Herein, we report a simple, phosphine- and thiol-free, effective hot-injection synthesis of colloidal 2D CuSe nanocrystals. This method uses non-coordinating 1-octadecene (ODE) as the solvent and oleylamine as both the reductant and ligand, thereby completely avoiding the use of toxic, pyrophoric, and expensive alkylphosphines, as well as dodecanethiol. Thiol ligands can reduce Se, introduce sulphur impurities or promote Cu–S side reactions, whereas thiol-free conditions avoid such competing pathways and enable cleaner growth of pure klockmannite CuSe[37] . The resulting nanocrystals form micron-sized nanosheets (NSs) in a wide lateral size range up to 4 μm, and uniform triangular nanoplatelets (NPLs) in the size range of 12–25 nm. We also discuss the steady-state and transient optical properties of our NCs, focusing on the influence of shape on plasmon resonances and absorption spectra, supported by advanced complex scaled discrete dipole approximation (CSDDA) simulations.

## Materials and Methods

### *Chemicals*
Unless stated otherwise all chemicals were ordered from Sigma-Aldrich and used without any further purification. Copper (I) iodide (CuI, 99.5%), selenium powder (Se, -100 mesh, 99.99%), oleylamine (OLA, 70%), octadecene (ODE, 90%), toluene (99.5%), hexane (VWR, 95%), isopropanol (VWR, 99.7%) and acetone (Th. Geyer, 99%).

*Synthesis of CuSe NCs*
CuSe NCs were synthesized using a hot injection method under an Argon atmosphere using a Schlenk-line setup. Initially, 2 mmol to 4 mmol Se was added to a mixture of OLA and ODE with a 1:3 ratio (8 – 16 mL) in a three-neck flask, and that mixture was heated at 200°C under an argon atmosphere till the Se was completely dissolved. Afterwards, the temperature was stabilized at 70°C and the solution was degassed by applying vacuum for 1 hour. Then the flask was filled with argon and gradually heated up to the required synthesis temperature. Further, 3mL of a previously degassed CuI/OLA solution (0.05 to 0.2 mmol of CuI) was hot injected at 200-220°C. Aliquots were taken during the reaction at different time scales to study the formation mechanism and growth of the NCs. The reaction mixture was refluxed between 5-10 minutes and then cooled down to room temperature by quenching it with a cold-water bath. For purification, the crude CuSe NCs were precipitated by adding an acetone/isopropanol mixture and then extracted by centrifugation. The precipitated product was resuspended in hexane and the washing procedure was repeated twice. The final purified NCs were dispersed in hexane for further characterization use.

*Characterization of the CuSe NCs*
X-ray diffraction (XRD) measurements were performed on a Malvern-Panalytical Aeris System with an X-ray wavelength of 0.154 nm from the Cu-K$\alpha$1 line. The samples were prepared by drop-casting the suspended NCs on a low background silicon wafer substrate (<911> or <711> cut). TEM samples were prepared by diluting the suspension with hexane followed by drop-casting it on a copper grid with a carbon film. Standard TEM images and selected area electron diffraction (SAED) patterns were acquired on a Thermofisher Talos-L120C microscope with a thermal emitter operated at an acceleration voltage of 120kV. High-resolution STEM images were acquired on a Jeol ARM20CF NeoARM microscope at 200 kV acceleration voltage. UV/Vis/NIR absorption spectra were obtained using a quartz cuvette with a Lambda 1050+ spectrophotometer from PerkinElmer equipped with a 150 mm-large integration sphere. Transient absorption (TA) spectra were recorded with ~100 fs resolution using a pump-probe setup with a white light continuum for probing and a noncollinear optical-parametric amplifier (NOPA) as excitation source (460 nm, ~70 fs pulses)[38]. Both were pumped by a Ti:sapphire laser system (CPA 2001, Clark MRX) at 775nm and a repetition rate of 1kHz. The probe was dispersed by a prism and detected via a photodiode array. To minimize scattering effects, the pump and probe pulses were cross-polarized to filter out the scattered pump light with a linear polariser in front of the detector. Samples in hexane were drop-cast onto quartz substrates resulting in optical densities of ~0.15-0.25 at 460 nm. Raman spectra were obtained using a self-built micro-Raman spectroscopy setup on samples drop-cast onto silicon wafer pieces with a top $SiO_2$ layer. For the Raman measurements, a linearly polarized, diode-pumped solid-state laser with a wavelength of 532 nm was focused onto the samples using a 50x microscope objective to a spot size of about 1 μm. The backscattered light was collected by the same objective. The elastically scattered light was filtered out using a set of three reflective Bragg grating notch filters (Optigrate), giving access to Stokes- and Anti-Stokes Raman signals. The inelastically scattered light was further analysed using a linear polarizer to yield cross-polarized detection and detected using a grating spectrometer and a Peltier-cooled charge-coupled-device camera.

*Band structure and CSDDA++ calculations*
First-principles computations were performed using the Questaal suite[39] for the band structure and optical response of CuSe. A modified complex-scaled discrete dipole approximation[40] (CSDDA++) was applied to simulate field excitations within CuSe NCs, using the obtained dielectric functions. Our version enhances conventional DDA[41] by optimizing the iterative scheme in the complex plane, achieving faster convergence. Numerical simulations were conducted on the Zarquon Cluster (224 CPU threads) with cubic dipoles of 1.5 Å. For a 20 nm NC, over one million dipoles were used, covering a volume of 3456 nm³. Simulation times ranged from a few hours for the dielectric domain to 26–30 hours for the hyperbolic domain. The resulting intensity-normalised absorption spectra are presented in the SI for various shapes and field incidences, following ref [42].

**Results and Discussion**

*Formation and morphology of the nanocrystals*
The quasi-2D CuSe nanostructures were synthesized using the hot-injection approach followed by size-selective precipitation. A schematic representation of the synthesis process is shown in Figure S1, and a detailed description of the synthesis technique is provided in the Experimental section. CuI was used as the $Cu^+$ precursor to stabilise copper via soft Lewis acid–base coordination with iodide, thereby moderating its reactivity and favouring stoichiometric $Cu_1Se$ [43,44]. In a typical synthesis, CuI in OLA (0.2 mmol) was injected into a Se solution (4 mmol) in ODE and OLA at 200 °C without additional ligands. The combination of OLA and a selenium-rich environment is the key factor directing the reaction toward the klockmannite CuSe phase and suppressing formation of non-stoichiometric $Cu_{2-x}Se$. This ligand–anion environment moderates $Cu^+$ reactivity and stabilises the 2D phase.

The precipitate consisted of nanosheets (NSs) with a nonuniform size distribution, with lateral sizes of 0.2–4 μm and thicknesses of 5–15 nm as presented in Fig. 1(a,b). High-resolution TEM (HRTEM) analysis and SAED revealed that these NSs have a hexagonal klockmannite crystal phase with $a=b=3.94$ Å and $c=17.25$ Å (Figure 1(c), (d)). The SAED pattern is indexed along the [001] zone axis, confirming the single-crystalline nature of the NSs. XRD further verifies phase purity, with the diffraction peaks at 2θ = 10.1°, 27.8°, 31.0°, 41.9°, and 53.1° corresponding to the (002), (102), (006), (008), and (0010) planes of klockmannite CuSe. The reduced number of peaks and the strong (006) reflection arise from texture effects associated with preferential [001] orientation on the Si substrate, consistent with SAED observations and the simulated XRD pattern of a [001]-oriented klockmannite crystal (Fig. S2).

When gradually decreasing the Cu concentration in the synthesis, hexagonal NPLs (Cu:Se = 0.1:4 mmol) and later faceted triangular NPLs (Cu:Se = 0.05:4 mmol) with an average altitude size of 16 nm and thickness of ≤5 nm are formed together with micron-sized NSs. Their 2D morphology is evident from homogeneous TEM contrast, HRTEM fringes, lateral overlap and stacking behaviour. These NPLs were separated using size-selective precipitation and filtration procedure (Fig. S1). After quenching the reaction, the crude product was centrifuged to collect the supernatant (containing NPLs and some small NSs) and the supernatant was purified via repeated low-RPM (1500) centrifugation. Additional separation was achieved using 200 nm PTFE filters, which improved the separation and purity of the small NPLs. Across repeated syntheses, triangular NPLs were obtained with high reproducibility, comprising ~50-60% of the isolated nanocrystal fraction. Their size-distribution histograms (Fig. S1-b) confirm relatively narrow dispersity (standard deviation of 12%) and support uniformity.

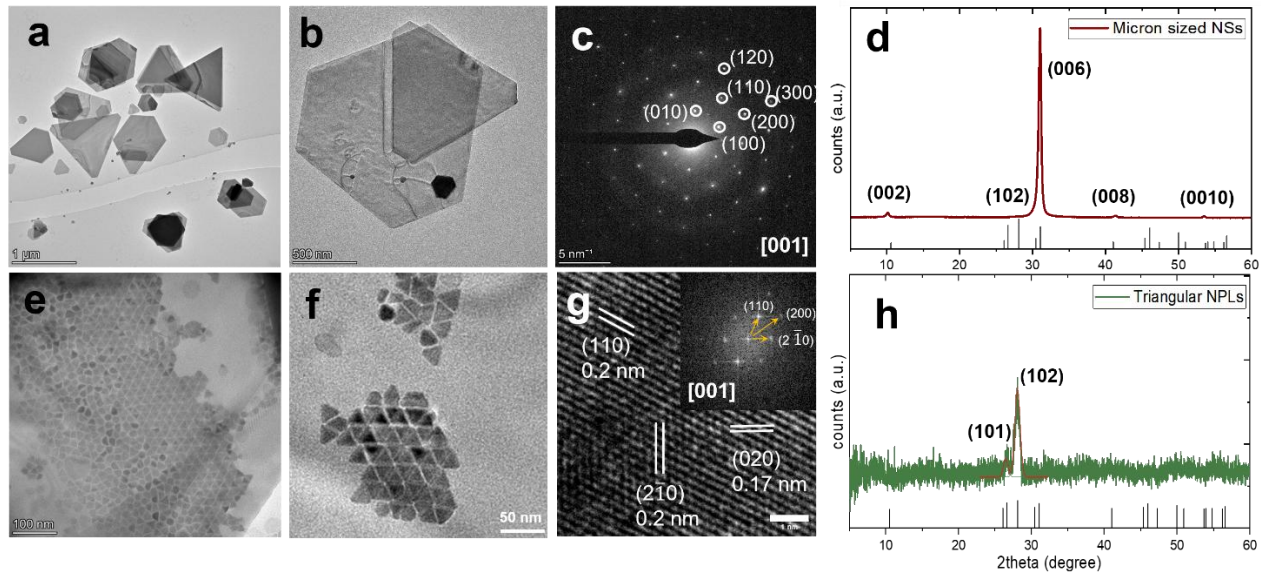

**Fig. 1** Structural characterization of the klockmannite CuSe nanostructures. (a,b)TEM images of micron sized NSs (c) SAED of single NS along the [001] zone axis (d) respective powder XRD of NSs. (e,f) TEM images of triangular NPLs (g) HRTEM of single NPL with FFT along the [001] axis (h)respective powder XRD of NPLS with fitted profile for (101) and (102) reflections. XRD reference PDF card No. 34-0171 of klockmannite CuSe phase was used.

Hexagonal NPLs crystallize in the berzelianite phase with some contribution from large klockmannite NSs (Fig. S3-1), whereas triangular NPLs exhibit a pure klockmannite structure, as confirmed by XRD (Fig. 1h) and HRTEM (Fig. 1g). In triangular NPLs, HRTEM lattice fringes correspond to d-spacings of ~0.2 nm for the (110) planes and 0.17 nm for the (020) planes, indexed along the [001] zone axis. Formation of triangular klockmannite structures is highly sensitive to precursor mixing and the local redox environment, often leading to mixed or hexa-triangular morphologies in the literature[45,46]. In our case, the triangular shape originates from the anion-rich environment created by excess selenium, which ensures Se-terminated early nuclei, suppresses Cu-rich intermediates and accelerates nucleation in the kinetic regime.[47] High local concentrations of reactive Se-OLA species favour lateral 2D growth, enabling thin triangular plates with large surface-to-volume ratios. During growth, OLA acts as both reductant and ligand, passivating Cu cations due to the nucleophilic nature of the amine group. DFT studies for copper sulfide[48] show higher amine adsorption energies on Cu-rich (100) facets than on (110), supporting by analogy the formation of triangular and hexagonal platelets with (100)-like edge facets in CuSe.

In addition to the specific choice of surfactants, reaction temperature plays a critical role in controlling the phase of CuSe NCs. For instance, syntheses performed below 200°C yielded berzelianite phase ($Cu_{2-x}Se$) NCs with cubic morphology and uniform size distribution[49] (Fig. S3-2), while syntheses conducted over 230°C produced a berzelianite phase ($Cu_{1.8}Se$) mixture with micron-sized klockmannite NSs (Fig. S3-3). The formation of various $Cu_{2-x}Se$ phases merits further investigation, as this system can crystallize into multiple closely related structures with small differences in formation energy[33], each capable of accommodating a broad range of copper stoichiometries[50]. In our study, we find that only a narrow range of experimental conditions promotes the formation of the "metastable" klockmannite phase of copper selenide.

### Metallicity of CuSe NSs

CuSe is known as a semiconductor with high p-type conductivity, demonstrated in both large-scale films[51, 52] and, more recently, in films of nanoparticles or nanosheets[6,19,31,33,53]. The origin of its metallic character can be attributed to the mixed Cu/Se valence states (+1, +2 / 0, -2), arising from strong Cu $d$ – Se $p$ hybridization, and to the presence of delocalized holes in the valence band, analogous to covellite copper monosulfide.

While films of CuSe have been extensively studied, individual colloidal NSs had not previously been contacted and electrically characterized. To investigate this, we contacted several individual NSs using e-beam lithography and observed nearly linear I–V curves with metallic behaviour, i.e. a positive temperature coefficient of resistivity (using cryo-probe station), which is characteristic for a metallic behavior. The values of the specific conductivity were 645.2 $(\Omega \cdot cm)^{-1}$ at 293 K and 1266.5 $(\Omega \cdot cm)^{-1}$ at 5 K (Fig S4), consistent with previously reported values for continuous films and nanoparticles. Our measured room temperature conductivity is approximately half of that reported for thin films in Refs.[51,52]. This discrepancy can be attributed primarily to the residual L-type surface ligand oleylamine, whose nucleophilic and electron-donating tether group[54] can attract and partially block quasi-free holes at the nanocrystal surface. In addition, contact resistance induced by residual ligands, enhanced charge scattering at the surfaces of thin NSs, and a reduced channel width likely contribute to the reduction in conductivity. Nevertheless, the measured values remain competitive for various applications and are comparable to those of highly conductive polymers. More detailed insight about the electrical properties of CuSe NSs will be given elsewhere.

### Optical steady-state properties of CuSe NCs in the visible and NIR regions.

Along with their high electrical conductivity and controlled morphology, the optical properties of CuSe deserve detailed examination. Fig. 2 shows typical optical absorption spectra of CuSe NCs suspended in hexane and measured in a quartz cuvette. A broad absorption feature appears between 400-600 nm and extends as a tail into the ultraviolet. The optical band gaps of ~2.5 eV for NPLs and 2.2 eV for NSs were estimated using Tauc linearisation by plotting of $(\alpha h\nu)^2$ versus photon energy ($h\nu$) (shown in Fig. 2b,d, where $\alpha$ is the absorption coefficient) assuming a direct band gap[55]. The blue-shifted band gap of the NPLs relative to bulk CuSe (2.2 eV)[8], likely reflects a combination of the Moss–Burstein and quantum confinement effect[20].

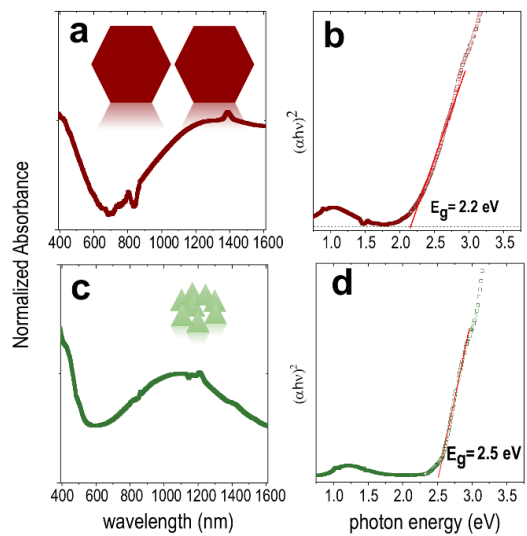

Fig. 2 (a,c) Absorption spectra of nanosheets in red and triangular nanoplatelets in green. (b,d) Tauc plots of the respective spectra considering CuSe as direct band-gap semiconductor.

Many cation-deficient copper chalcogenides are degenerately doped semiconductors, tolerating a large number of ion vacancies and thus high charge carrier concentrations[56,57]. As discussed earlier, klockmannite CuSe supports a high density of quasi-free holes due to its specific crystal structure, similar to CuS as established through theoretical and particularly DFT studies[58–61]. This allows CuSe NCs to sustain high intensity LSPRs, particularly in the NIR region[62] while shape and size tunability provide extensive control over these resonances. Freshly prepared CuSe NSs exhibit a broad, nonsymmetric plasmon band peaking at around 1400 nm, whereas smaller triangular NPLs (average size ~12–25 nm) display a narrower peak at ~1100 nm. Similar features are found for hexagonal NPLs in the cubic (isotropic) berzelianite phase (Fig. S5), which also exhibit strong localized surface plasmon bands in the NIR due to copper vacancies. The redshift of the bands in NSs with respect to NPLs is attributed to the larger lateral size resulting from slower and more balanced growth. Sharp features in the spectrum in the NIR originate from the measurement (solvent) artefacts.

Since both triangular NPLs and large NSs crystallise in the klockmannite phase, it is important to analyse their plasmonic properties in the NIR while accounting for anisotropy. Simulating LSPRs in irregular or complex geometries requires numerical methods, with the dielectric permittivity dispersion as a key input. This dispersion can be approximated within the Drude–Sommerfeld theory[48,63]. An essential drawback of this approach for CuSe is the necessity of several fitting parameters (plasma frequency, damping constant, bound electrons polarizability parameter $\varepsilon_\infty$), the absence of fundamental absorption and – more importantly – the absence of anisotropy. A more rigorous alternative is to obtain the band structure and construct the full permittivity tensor.

*DFT Calculations*
The band structure of CuSe was calculated using the quasiparticle self-consistent GW (QSGW) method, and the permittivity along two crystallographic directions (in-plane and out-of-plane) was derived using the random phase approximation (RPA) following the approach of Zayats et al.[42]. These optical parameters were then used in CSDDA simulations to model the near-field distributions around CuSe nanotriangles.

Figure 3a shows the CuSe band structure obtained from the QSGW+RPA method, with the Fermi level set to 0 eV (actual value: 0.479 eV). The Fermi level lies within the valence band, confirming that CuSe is a p-type semiconductor with a direct band gap of 0.257 eV at the Γ-point. The valence band originates from hybridisation between Cu d-orbitals and Se p-orbitals, while the conduction band derives from their respective s-states. Hole dispersion is minimal along the Γ–A ([001]) direction but significantly larger along Γ–K (in-plane), highlighting the strong electronic anisotropy of CuSe. Likewise, the conduction band shows its highest dispersion along Γ–K. The effective masses of the holes, extracted from the band dispersion, are 0.86·me in the xy-plane and 1.05·me along z; inclusion of spin–orbit coupling further increases this discrepancy.

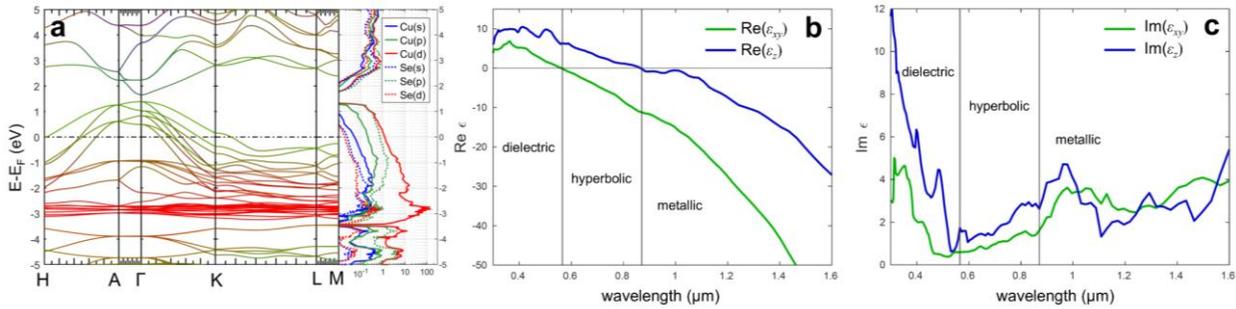

**Fig. 3** Electronic band structure and optical survey of CuSe using the QSGW-RPA approximation: (a) band structure with color coding: blue - *s* orbitals of Cu and Se, green - *p* orbitals and red - *d* orbitals, inset – projected density of states in CuSe, the Fermi level is set to 0 eV; (b) real part and (c) imaginary part of the dielectric function along xy (in-plane) and z direction (out-of-plane, i.e. [001] direction).

The dielectric function $\varepsilon(\omega)$ of CuSe was calculated over a broad energy range from 0.8 up to 4 eV (Fig. 3b, c). In Fig. 3b, the real part of the dielectric function shows that CuSe exhibits a dielectric response in the Vis range below 550 nm where Re($\varepsilon(\omega)$) is positive and Im($\varepsilon$) high (Fig. 3c). In NIR range above 900 nm, both out-of-plane and in-plane components of Re($\varepsilon(\omega)$) are negative and the light propagation is forbidden, making the material metallic. In the spectral range from nearly 550 to 880 nm, a hyperbolic optical response exists, where the permittivity in the xy-plane (Re[$\varepsilon_{xx}$], Re[$\varepsilon_{yy}$]) has an opposite sign to that along the z-axis (Re[$\varepsilon_{zz}$]). This hyperbolicity extends partially into the visible and the NIR regions. The existence of the hyperbolic range in the vicinity of the plasma frequency is expected for anisotropic materials since the effective masses are different for in- and out-of-plane directions. From classical Drude theory it is known that the plasma frequency is a parameter which defines the zero-crossing point of the permittivity. After $\varepsilon$ becomes negative, propagation of light in the medium is shielded by the plasma oscillation and the medium becomes strongly reflective. Being a function of the charge carrier concentration $N$ and their effective mass $m^*$ ($\omega_p = \left[\frac{eN}{\varepsilon_0 m^*}\right]^{\frac{1}{2}}$), the plasma frequency becomes a tensor for anisotropic media allowing for a hyperbolic domain. We note that the absolute position of the zero-crossing points of the permittivity tensor based on DFT calculations is an approximation, however, the emergence of the hyperbolic domain is a principal result with higher degree of plausibility.

With the obtained dielectric function, the plasmonic responses to incident fields were further studied in detail for CuSe NCs. The intensity-normalised absorption spectra calculated from CSDDA++ (see details in SI) for triangular NPLs are shown in Fig. 4 and in Fig. S6 for other possible shapes for comparison. The edge length of the NPLs in the simulations is 19 nm and the thickness *is* 5 nm.

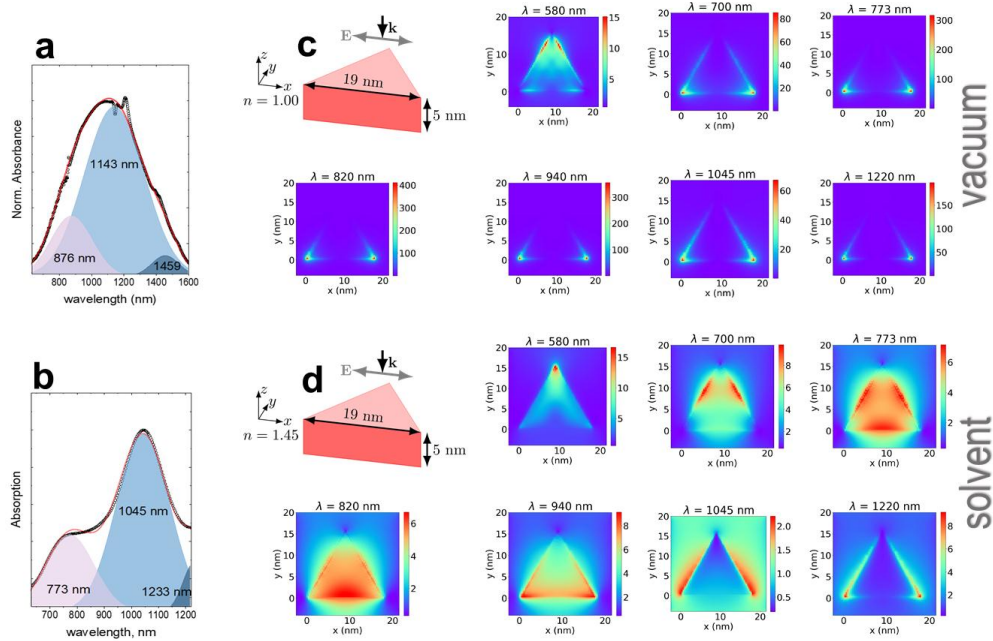

Fig. 4 (a) Experimental absorbance of the triangular CuSe NPLs in solution (n≈1.45) and (b) simulated orientation-averaged absorption cross-section (nm$^2$) of the triangular CuSe NPLs. (c, d) Predicted spatial electric field intensity distributions 0.1 nm above the surface of the NPL for different media around the NPL. (c) vacuum (no solvent); (d) with a solvent (n = 1.45).

During the simulations, we averaged the absorption spectra for changing polarization with respect to the plane of the nanoprism (idealised model geometry for triangular NPLs) surface in three directions (so called "orientational averaging"). First, we consider the collective spectrum of triangular NPLs experimentally measured in hexane (Fig. 4a). For qualitative understanding, we fitted the absorbance curve with three Gaussian functions. A similar procedure has been performed for the calculated spectrum in Fig. 4b. In the measured spectrum, the main contribution comes from the central peak at ~1140 nm, attributed to the strongest LSPR mode which is dipolar. Side peaks are needed to compensate for the residual absorption and asymmetry characteristic of the triangular NPLs. The long-wavelength component is likely coming from a size distribution shifted towards slightly larger particles, while the short-wavelength contribution likely arises from quadrupolar resonance or out-of-plane modes. Considering the anisotropy of CuSe in that spectral range and the predicted hyperbolicity, the out-of-plane mode can be ruled out. In the calculated spectrum (Fig. 4b), a dominant NIR LSPR appears at 1045 nm, accompanied by a higher-energy shoulder near 773 nm. The sensitivity of this shoulder to nanocrystal orientation is illustrated in Figs. S7–S9. The main peak in the measured spectrum (~1143 nm) aligns closely with the simulated peak (~1045 nm), red-shifted by only ~0.1 eV despite no fitting parameters. The high-energy shoulder in the experiment is weaker and slightly more red-shifted (~0.2 eV) than in the simulations. Additional spectra (Figs. S6 and S8) reveal that this short-wavelength feature depends strongly on corner geometry and disappears for disk-like platelets. This suggests that the vertices of the experimental triangular NPLs are not perfectly sharp but slightly rounded. Indeed, the corner curvature radius is estimated to be 2.8–4 nm for a triangular NPLs with an average the edge size (Fig. S10). Among further factors that can broaden, shift, or quench resonances chemical interface damping can be mentioned due to the presence of a monolayer of oleylamine on the surface[64]. Finally, collective ensemble effects such as size distribution, NC stacking, orientational disorder, light scattering, and additional LSPR damping from surface traps may influence the measured spectra and are not yet included in our simulations.

In Fig. 4 (c, d), intensity profiles above the surface of the triangular NPL at different wavelengths (in cgs units, i.e. statV/cm$^2$) are shown for different surroundings – vacuum and solvent (corresponding to the experiment $n$ = 1.45) and solvent+substrate (SiO$_2$) in Fig. S11 for comparison. For the NC in vacuum, the resulting plasmonic excitations are tightly confined to the corners of the NC along the whole spectrum (Fig. 4c). These spatial field distributions are similar to what one would expect in an isotropic medium. The resulting field enhancement is high. When the solvent is introduced (Fig. 4d), wavelengths beyond 900 nm—corresponding to the fully metallic regime—produce dipolar-like field distributions localised near the corners (at λ = 1045 and 1220 nm), characteristic of the LSPRs commonly observed in metals[65] and even for copper sulfide[42,48]. In the hyperbolic domain, however, a significant deviation from the spatial intensity distribution of (c) is noted. Specifically, in the hyperbolic domain from 890 nm to 550 nm, the plasmonic excitation spreads beyond the vertices. Near the epsilon-near-zero wavelength (~820 nm), an "envelope" excitation mode forms, covering nearly the entire surface of the NC, with notable enhancement along the edges. This hybrid mode combines quadrupolar features with an envelope-type excitation. The predicted effect of the solvent + an infinite-plate substrate can be found in Fig. S9b. The addition of the infinite plate has a subtle but distinct effect on the plasmonic excitation at 820 nm from configuration (d) in Fig. 4, suppressing the edge excitations. The plasmonic mode is now tightly confined within the edges of the NC in the hyperbolic domain. Simulations of absorption spectra for other shapes such as truncated triangles, hexagons and disks are also shown in the SI for comparison. We assume that in the hyperbolic spectral domain, a large range of evanescent modes on the surface of the nanocrystals in a solvent is allowed, and the field distribution gradually progresses towards the "envelope" mode on the entire surface of the NC as a result of a combination of propagating and evanescent fields. To disentangle the roles of shape and intrinsic crystal anisotropy, we simulated an isotropic "model" CuSe nanoprism (Fig. S12). These simulations reveal that shape anisotropy determines the spectral shifts, hotspot locations, and asymmetric spectral broadening, whereas the intrinsic material anisotropy is solely responsible for the envelope-type modes, which appear only in the anisotropic nanocrystal.

***Ultrafast dynamics of charge carriers in the excited state***

Along with their unique stationary optical properties, the ultrafast behavior of copper chalcogenides has also gained increasing attention in recent years. Relaxation and recombination dynamics are particularly crucial for these nanocrystals in energy-transfer related applications. According to available studies on CuS [17,66–71], these materials differ from metals and semiconductors in their ultrafast relaxation dynamics due to their hybrid properties combining the fundamental band gap, specific concentration of quasi-free charge carriers (very high for semiconductors and low in comparison to metals), surface phenomena and geometrical constraints imposed by synthesis conditions. In this regard, possible applications include but are not limited to hot-carrier extraction, ultrafast switching and photodetection.

To explore these properties further, we performed ultrafast pump-probe transient absorption (TA) measurements on anisotropic klockmannite CuSe NCs, exciting the samples in the fundamental absorption region at 460 nm and tracking the induced absorption changes in the visible spectral range. As shown in Fig. 5a and 5b, the TA spectra are dominated by a broad excited-state absorption (ESA) that overlaps with the ground-state bleach (GSB). The spectral shape and position of the GSB, indicated in grey in Fig. 5c and 5d, is obtained from the respective steady-state spectra of the NCs presented in previous sections.

To analyse the TA data $\Delta OD(\lambda, t)$, we performed a global lifetime analysis (GLA), fitting the transient spectra with a sum of exponential decay components using the function:

$$\Delta OD(\lambda, t) = \sum_i A_i(\lambda) \cdot \exp\left(-\frac{t}{\tau_i}\right)$$

The $i$-th decay component is characterized by its time constant $\tau_i$ and the decay associated amplitude spectrum (DAS) $A_i(\lambda)$. The obtained time constants are summarized in Table 1, and the corresponding DASs for NSs and NPLs are shown in Fig. 5c and 5d, respectively. For both samples, the observed dynamics consists of three components: a fast, a medium-fast, and a slow one. We interpret these results assuming that CuSe behaves similarly to the covellite CuS, which exhibits comparable mixed-valence character, plasmonic response, and ultrafast relaxation characteristics.

**Table 1:** Time constants obtained by GLA of the TA data as well as frequencies and damping constants of the observed oscillations.

|  | $\tau_1$ [ps] | $\tau_2$ [ps] | $\tau_3$ [ns] | $f$ [THz] | $\tau_d$ [ps] |
|---|---|---|---|---|---|
| **NSs** | 1.3 | 99 | > 2 | 7.5 | 0.5 |
| **NPLs** | 0.9 | 92 | > 2 | 7.8 | 0.7 |

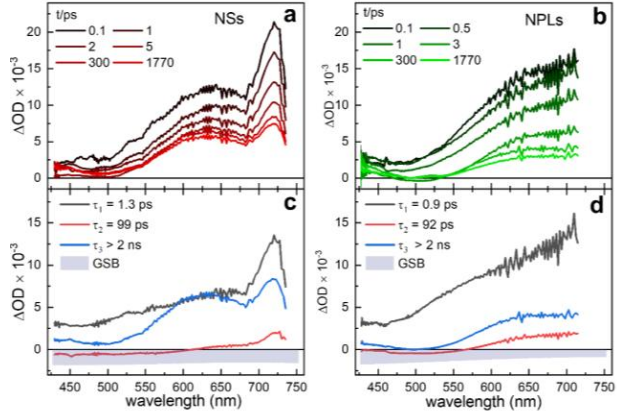

**Fig. 5** (a, b) TA spectra of NSs and NPLs at different times t after optical excitation at 460 nm. (c, d) DASs for NSs and NPLs obtained by the GLA and labelled by the respective time constants. The dynamics is well described by a multi exponential decay with the time constants $\tau_1 = 1.3$ ps, $\tau_2 = 99$ ps and $\tau_3 > 2$ ns for NSs and the time constants $\tau_1 = 0.9$ ps, $\tau_2 = 92$ ps and $\tau_3 > 2$ ns for NPLs.

The shortest time constant $\tau_1$ is attributed to the hole trapping process[17,70,72], occurring slightly faster in NPLs ($\tau_1 = 0.9$ ps) compared to NSs ($\tau_1 = 1.3$ ps). The two longer time constants $\tau_2$ and $\tau_3$ reflect a combination of photocarrier recombination and the relaxation of trapped holes[17,70–72]. In both cases, the component with $\tau_2$ (99 ps for NSs and 92 ps for NPLs) contributes only weakly and reflects a slight narrowing of the ESA signal in the red spectral region and a minor increase in the blue spectral region. The final decay occurs on the nanosecond timescale and is described by the time component $\tau_3$. Due to the experimental setup, which is limited to a maximum delay of 1.8 ns, an exact value for $\tau_3$ cannot be determined. However, the data suggest that the decay in NSs is slightly slower than in NPLs. The tendency toward shorter time constants can be explained by the fact that smaller NPLs generally have higher defect densities and surface-to-volume ratios than NSs. Defects can serve as recombination centers, which accelerate the relaxation dynamics.[73,74]

Particularly interesting are the oscillatory features observed in the TA time traces during the first two picoseconds (Fig. 6a). The ultrafast excitation of hot holes and electrons is accompanied by the generation of coherent phonons[69,75]. To analyse these oscillations in more detail, the time traces in the wavelength range of 480–500 nm were averaged, and the first 2 ps were isolated. The underlying exponential decay was subtracted to extract the oscillatory components, and a fast Fourier transformation (FFT) was applied to the resulting residuals. The results, presented in figures 6b and 6c for NSs and NPLs, reveal a pronounced mode at 7.6 THz for NSs and at 7.8 THz for NPLs. These modes can be confirmed by the measured Raman spectra of the samples (Fig. 6 d–e), which show vibrational modes at about 260 cm$^{-1}$ assigned to the Se–Se stretching A$_{1g}$ mode[76]. The oscillatory features can also be reproduced using a model containing two exponential decay terms and a damped oscillatory component. The oscillation frequencies extracted from the fits are 7.8 and 7.5 THz (Table 1; see SI for details), are in good agreement with the FFT analysis. Interestingly, the NPLs exhibit a slightly larger damping constant compared to the NSs.

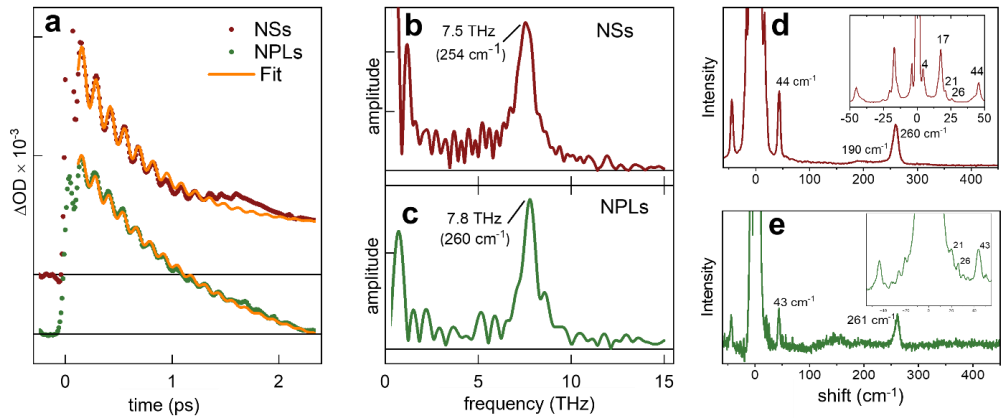

**Fig. 6** Oscillatory TA features and corresponding Raman spectra; (a) TA time traces averaged from 480 nm – 500 nm from NSs (red) and NPLs (green dots) with corresponding fits (orange curve). (b and c) FFT spectra of the residuals resulting from subtracting the exponential contributions from the averaged time traces for the samples NSs and NPLs. (d and e) Raman spectra of the CuSe NCs on a Si/SiO$_x$ substrate in full and reduced spectral range (zoomed) for NSs (red curve) and NPLs (green curve).

The phase of the oscillation can provide insight into the mechanism of coherent phonon generation[77]. However, the phase values obtained from the fit are not sufficiently precise, as the time-zero of our measurements is accurate only to ±30 fs, corresponding to approximately a quarter of an oscillation period. We attribute the formation of the coherent phonons to the displacive excitation of coherent phonons (DECP) mechanism[69], in which photoexcitation induces a sudden shift in the equilibrium geometry, leading to coherent lattice vibrations[77]. An alternative mechanism would be impulsive stimulated Raman scattering (ISRS): DECP produces coherent vibrations in the electronically excited state, whereas ISRS generates them in the ground state. Evidence supporting DECP arises from the fact that the oscillatory features in the TA data exhibit slightly lower frequencies than those obtained from Raman spectroscopy, which probes the ground state. Because bonds in the excited state are typically weaker, the corresponding vibrational frequencies are expected to be lower, consistent with our observations. We further averaged the FFT data across different spectral regions. One range spans the absorption minimum between the stationary UV absorption and the plasmon band, while the other covers the adjacent flank of the broad UV absorption. Our reasoning was that if the oscillations originated from ground-state vibrations, the stationary absorption spectrum should shift periodically with the vibrational frequency. In that scenario, the FFT amplitude would be strong on the spectral flank and weak in the absorption minimum. However, this behaviour is not observed; instead, the FFT amplitudes are comparable across all analysed regions (Fig. S13). This provides additional evidence that the observed wave packets are generated in the excited state.

Lastly, considering the Raman spectra, we note, besides the already discussed Se–Se stretch mode at 260 cm$^{-1}$, a weak mode at 190 cm$^{-1}$ that likely corresponds to a Cu–Se stretch vibration, similar to that reported for CuS[78]. Additional lower-frequency features, clearly identifiable on both the Stokes and anti-Stokes sides, include modes at ±44 cm$^{-1}$, ±26 cm$^{-1}$, ±21 cm$^{-1}$, ±17 cm$^{-1}$ for NSs, and ±43.5 cm$^{-1}$, ±26 cm$^{-1}$, ±21 cm$^{-1}$ for NPLs. The ±4 cm$^{-1}$ mode should be assigned to Brillouin scattering of the LA mode of the Si/SiO$_x$ substrate[79]. Features at 17 cm$^{-1}$ and 45 cm$^{-1}$ have been reported by Ishii et al.[76] and mentioned by Peiris et al.[36] in a CuSe compression study (though without interpretation). However, the remaining modes likely represent new low-frequency, unassigned Raman features and might correspond to rigid-layer shear and breathing modes, revealing the distinct (supposedly soft) interlayer coupling[73] in the quasi-layered structure of CuSe. The detection of coherent lattice oscillations together with the corresponding Raman signatures provides strong evidence for the klockmannite phase in both the CuSe NSs and triangular NPLs, and confirms their pronounced crystal anisotropy. While the excitation of coherent phonons in CuSe is only indirectly related to its anisotropy (the Se–Se A$_1$g stretch mode being aligned with the six-fold [001] axis of klockmannite), the emergence of the hyperbolic domain is a direct consequence of the intrinsic structural anisotropy. Thus, together with their plasmonic response in the NIR, anisotropic CuSe nanocrystals represent a promising material platform for advanced optoelectronic and photonic applications.

**Conclusions**

In this work, we presented a thiol-free colloidal synthesis of CuSe nanocrystals from nanoplatelets to micron-sized large nanosheets and demonstrated a certain shape control including hexagonal and triangular nanoplatelets. We investigated various properties of these structures, including metallic electrical conductivity at different temperatures, as well as their optical and structural characteristics. Using GW DFT and CSDDA simulations, we derived absorption spectra that align well with experimental data, showcasing the material's broad tunability – fundamental absorption in the visible (Vis) range, plasmonic activity in the near-infrared (NIR), and hyperbolicity across the Vis-NIR spectrum. This hyperbolic behaviour leads to unique field distributions, including enveloping surface modes, which result from the interaction between propagating and evanescent fields. This understanding opens up the possibility for engineering the optical response based on particle shape, orientation, and excitation wavelength. Additionally, we analysed the ultra-fast carrier relaxation and recombination dynamics in both nanosheets and nanoplatelets and observed a hot hole trapping process and the generation of coherent phonons, in agreement with Raman spectra. The Raman analysis also suggests a significant degree of anisotropy in CuSe, providing further insight into its unique properties and potential for advanced applications.

**Acknowledgements**

DFG is acknowledged by the authors for funding of SFB 1477 "Light-Matter Interactions at Interfaces", project number 441234705 as well as for financial support via the SPP 2102 "Light-controlled reactivity of metal complexes" (LO 714/11-2, project number 404479188).

Supporting Information

# Unlocking klockmannite: formation of colloidal quasi-2D CuSe nanocrystals and photo-physical properties arising from crystal anisotropy

Urvi Parekh[1], Nadiia Didukh[1], Samira Dabelstein[1], Ronja Piehler[1], Eugen Klein[1], Jivesh Kaushal[1], Tobias Korn[1], Stefan Lochbrunner[1,2], Christian Klinke[1,2], Stefan Scheel[1]*, Rostyslav Lesyuk[1,3]*

[1] Institute of Physics, University of Rostock, Albert-Einstein-Straße 23, 18059 Rostock, Germany
[2] Department "Life, Light & Matter", University of Rostock, Albert-Einstein-Straße 25, 18059 Rostock, Germany
[3] Pidstryhach Institute for applied problems of mechanics and mathematics of NAS of Ukraine, Naukowa str. 3b, 79060 Lviv, Ukraine

* corresponding authors: stefan.scheel@uni-rostock.de, rostyslav.lesyuk@uni-rostock.de

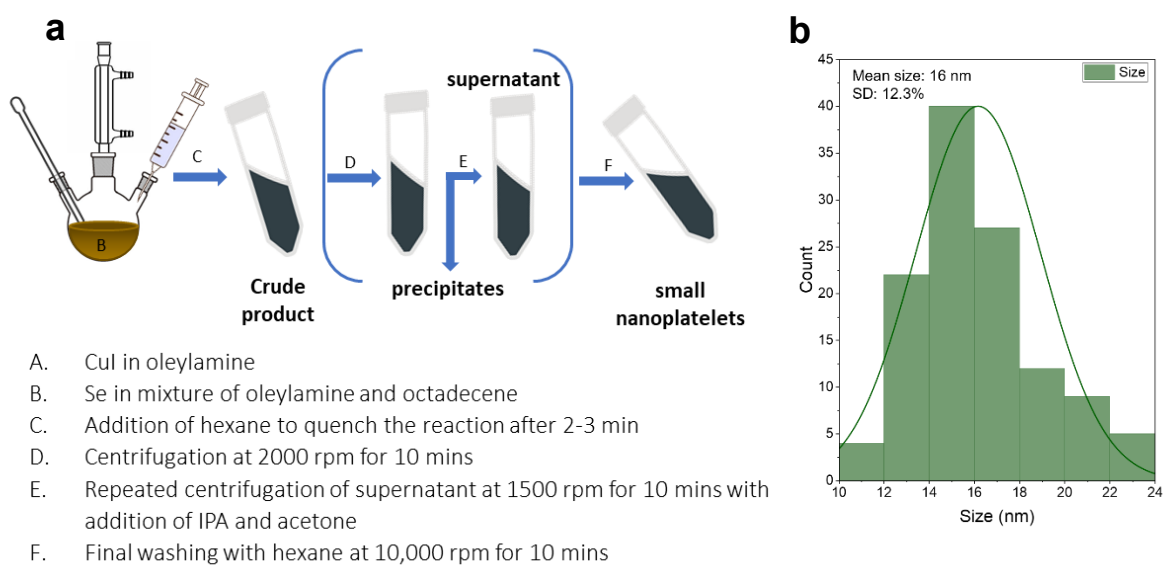

A. CuI in oleylamine
B. Se in mixture of oleylamine and octadecene
C. Addition of hexane to quench the reaction after 2-3 min
D. Centrifugation at 2000 rpm for 10 mins
E. Repeated centrifugation of supernatant at 1500 rpm for 10 mins with addition of IPA and acetone
F. Final washing with hexane at 10,000 rpm for 10 mins

**Fig. S1**. (a) Synthesis procedure of CuSe nanosheets with size-selective precipitation of nanoplatelets and (b) histogram showing the size distribution of nanoplatelets (altitude of the triangles).

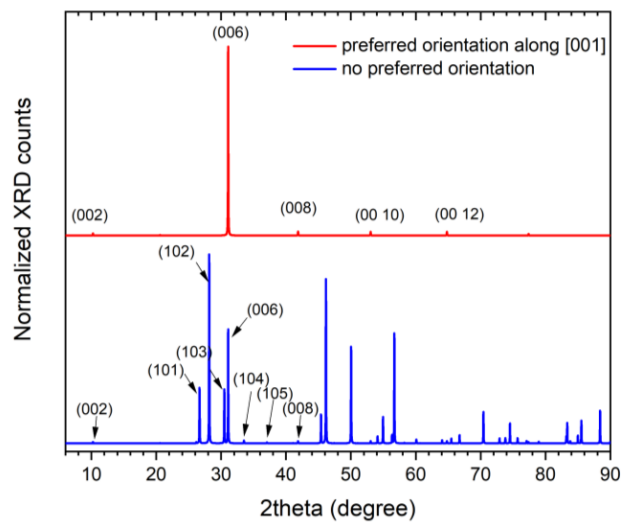

**Fig. S2**. Calculated XRD patterns of klockmannite CuSe crystal based on the crystallographic data of L.G. Berry [The crystal structure of covellite, CuS and klockmannite, CuSe American Mineralogist, 1954, 39, 504-509]. PowderCell 2.4 was used for the simulation[1].

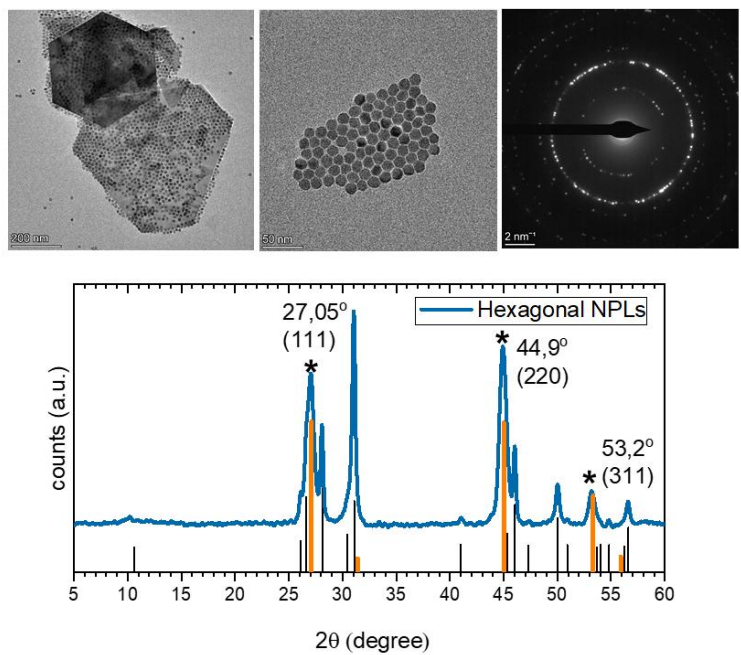

**Fig. S3-1.** Hexagonal NCs: TEM images, SAED and XRD pattern. Reflections marked with asterisk belong to the berzelianite phase (PDF card No. 01-088-2043).

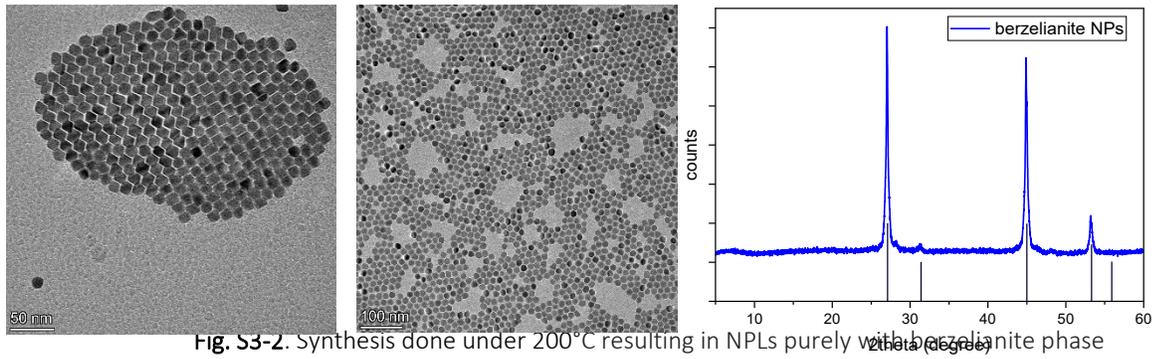

Fig. S3-2. Synthesis done under 200°C resulting in NPLs purely with berzelianite phase

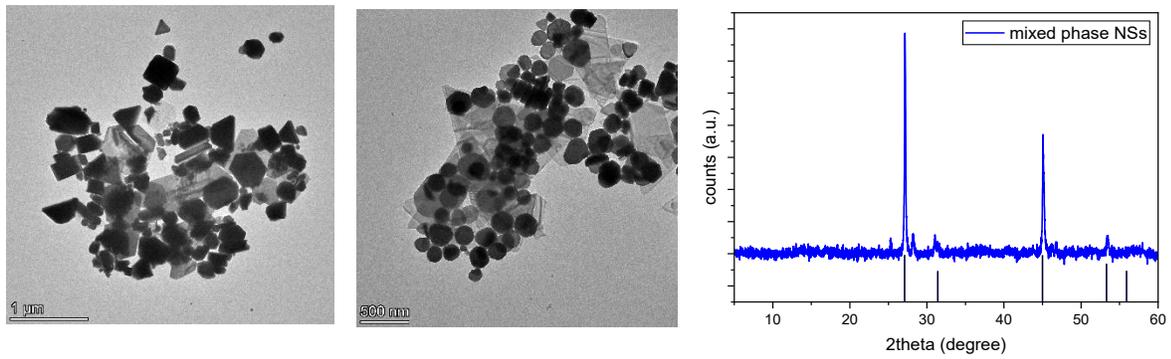

Fig. S3-3. Synthesis done over 220°C resulting in thick NSs of mixed phase predominantly containing berzelianite phase.

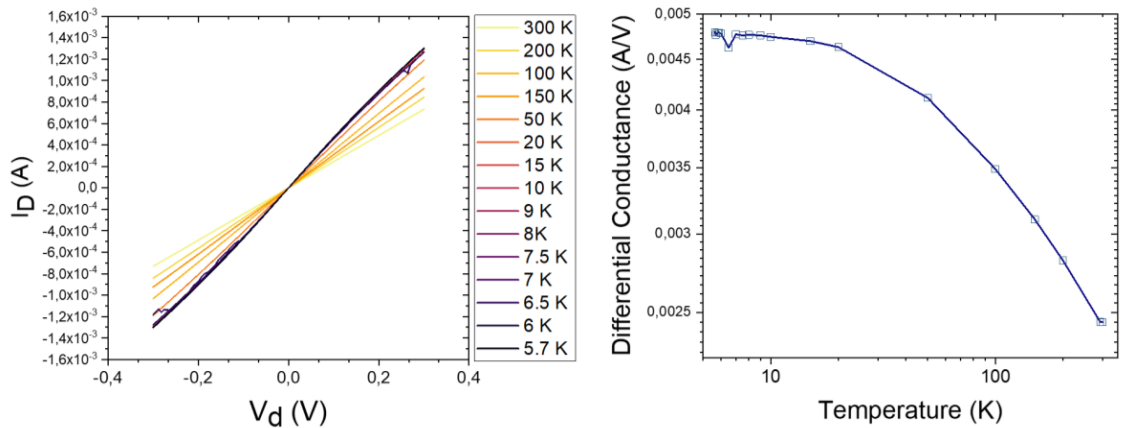

Fig. S4. Electrical transport through individual CuSe NSs.

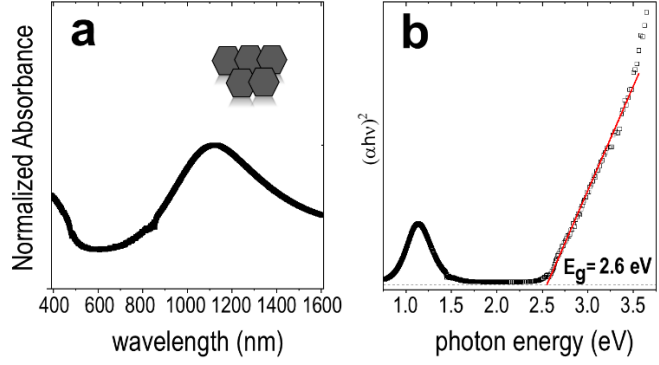

**Fig. S5**. Absorption spectra of hexagonal berzelianite NPLs with cubic crystal structure and respective Tauc plot. A strong plasmon band is present in the NIR.

**QSGW simulations**

To determine the optical properties of CuSe NCs in detail, quasi-particle self-consistent GW (QSGW) simulations and scattering calculations using the discrete-dipole approximation (DDA) were performed. To do so, we calculate the complex dielectric function $\varepsilon$ over a wide spectral range. We employ a method used previously by Zayats et al. [2] to extract the complex $\varepsilon$ of CuSe for two crystallographic directions (in-plane and out-of-plane) related to the strong anisotropy in this material.

First, the band-structure and optical response of bulk CuSe has been derived using the QSGW approach. Conventional density-functional theory (DFT) methods can deliver variety of structural, electronic, mechanical and thermodynamic properties [3–6] however they frequently lead to a band structure with the absence of a band gap. Along with time-dependent DFT based on solving a time-dependent Schrodinger equation with single-body electron density as a fundamental variable, the GW approximation (GWA)[7,8] presents another category of approximations that are based on many-body perturbation theory. It considers the exchange-correlation energy as a series of interactions involving multiple particles using one particle Green function (G) and screened Coulomb interaction (W) [9,10].

Quasi-single-particle GW (QSGW) gives a fairly accurate and qualitatively correct assessment of phenomena in most physical systems, in particular for CuSe, and together with random phase approximation (RPA) [2], it takes into account the plasmonic nature of the system.

**Complex-scaled discrete-dipole approximation (CSDDA) Method**

Here, we present the details of implementing the CSDDA algorithm. We first consider the standard recursive iteration at the heart of all DDA schemes. For the $n_{th}$ dipole, the local field is iteratively calculated using the algorithm [11]:

$$\boldsymbol{E}_{loc,n}^{(i+1)} = g^{(i)}\big(\boldsymbol{E}_{inc,n} + \boldsymbol{E}_{scatsum,n}^{(i)}\big) + (1 - g^{(i)})\boldsymbol{E}_{loc,n}^{(i)}$$

(1)

where $E_{loc,n}^{(i)}$ is the local field at the $n$th dipole, $E_{inc,n}$ is the incident field at the $n$th dipole, $E_{scatsum,n}^{(i)}$ is the scattered field at the $n$th dipole due to all other dipoles,

$$E_{scatsum,n}^{(i)} = \sum_{\substack{m=1 \\ m \neq n}}^{N} G_{nm} \cdot E_{loc,m}^{(i)}. \tag{2}$$

In Eq. (1), $g^{(i)}$ is an optimization parameter at iteration $i$. With CSDDA, the goal is to find a value of $g^{(i)}$ at every iteration step $i$, so that the relative error $R^{(i)}$,

$$R^{(i)} = \sqrt{\frac{\sum_{n=1}^{N} |E_{inc,n} + E_{scatsum,n}^{(i)} - E_{loc,n}^{(i+1)}|^2}{N}} \tag{3}$$

has the minimum possible value. This is done by choosing the value [12]: $g^{(i)} = -\frac{\sum_{n=1}^{N} X_n^{(i)*} \cdot Y_n^{(i)}}{\sum_{n=1}^{N} |X_n^{(i)}|^2} \tag{4}$ where $X_n^{(i)}$ is the net error between the scattered and local electric field vectors in Eq. (1),

$$X_n^{(i)} = E_{inc,n} + E_{scatsum,n}^{(i)} - E_{loc,n}^{(i)}, \tag{5}$$

and $Y_n^{(i)}$ tracks the net error between these two fields and the error propagated from every other dipole,

$$Y_n^{(i)} = -X_n^{(i)} + \sum_{\substack{m=1 \\ m \neq n}}^{N} G_{nm} \cdot X_m^{(i)}. \tag{6}$$

Convergence is achieved when both $X_n^{(i)}$ and $Y_n^{(i)}$ approach zero with each iterative step. For dielectric and/or isotropic material, it is usually true that the error propagated from every other dipole is smaller than the local error at the $n^{\text{th}}$ dipole. Thus the condition for $X_n^{(i)} \to 0$ and $Y_n^{(i)} \to 0$ is usually satisfied as we follow the optimization scheme Eq. (4) along each iteration step $i$. Then, as $g^{(i)} \to 0$, the CSDDA algorithm settles into a converged solution (i.e. $R^{(i+1)} \to 0$).

However, for anisotropic media, a new situation arises: we can still get $g^{(i)} \to 0$ but the relative error $R^{(i+1)}$ does not converge to zero. It can even be as high as 100 %. The reason for this paradox is the anisotropy of the medium, which allows for the following situation to emerge:

$$\sum_{n=1}^{N} X_n^{(i)*} \cdot Y_n^{(i)} \to 0. \tag{7}$$

Thus, from Eq. (4), we have still $g^{(i)} \to 0$, but there is no guarantee for the relative error $R^{(i+1)} \to 0$, and in fact it does not (Fig. *CSDDA-1*).

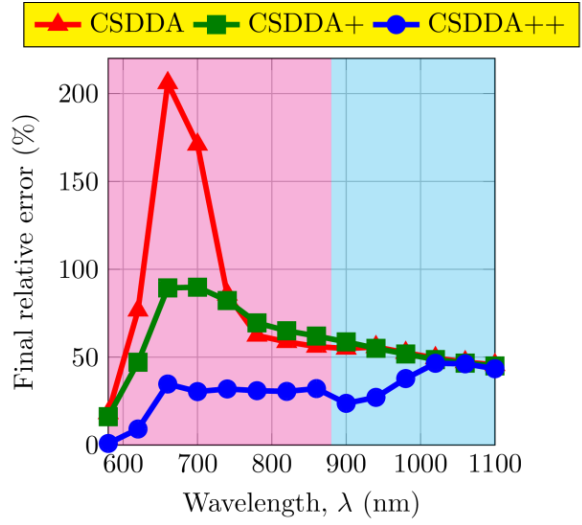

**Fig. *CSDDA-1*** Comparison of relative error around $g^{(i)} \to 0$, when the iteration scheme practically stagnates, for anisotropic CuSe bulk media: conventional CSDDA (red triangle), double complex CSDDA (CSDDA+) without taking into account error propagation effects (green squares), and CSDDA++ that takes into account both the error propagation and allows optimization over largest complex parameter space (blue circles). We clearly see the massive errors in conventional CSDDA approach for anisotropic media, especially in the hyperbolic domain, where the anisotropy is the most pronounced. The hyperbolic regime therefore needs the full CSDDA++ iteration scheme shown in Eq. (9). The errors in the metallic domain are primarily due to the high field enhancement at the tips of the triangle.

Expanding Eq. (7), we can understand the source of the large errors:

$$-|X_n^{(i)}|^2 + \sum_{\substack{m=1 \\ m \neq n}}^{N} X_n^{(i)*} \cdot G_{nm} \cdot X_m^{(i)} \to 0 \Rightarrow |X_n^{(i)}|^2 \simeq \sum_{\substack{m=1 \\ m \neq n}}^{N} X_n^{(i)*} \cdot G_{nm} \cdot X_m^{(i)}$$

(8)

or, in other words, the error propagated to dipole $n$ from every other dipole is on the same order of magnitude as the local error at dipole $n$. This systemic error accumulation in CSDDA thus needs to be countered.

This has to be done in two ways:

1. by acknowledging the impact of error propagating from every other dipole $m \neq n$, in the iteration scheme Eq. (1) itself, and
2. by allowing for an optimization scheme over a much larger complex parameter space than in Eq. (1).

Not including the error propagation effects while still expanding the complex parameter space (using two optimizing hyperparameters $g^{(i)}$ and $h^{(i)}$ in Eq. (1) instead than just one) is still not sufficient, cf.~the date shown with green squares vs. those in blue circles in Fig. *CSDDA-1*.

We thus use the updated CSDDA method, which we call CSDDA++ (expanding to second-order propagation and expanding the optimizing parameter space), to implement the generalized recursive iteration scheme,

$$\boldsymbol{E}_{loc,n}^{(i+1)} = \alpha^{(i)}\left(\boldsymbol{E}_{inc,n} + \boldsymbol{E}_{scatsum,n}^{(i)}\right) + \beta^{(i)}\boldsymbol{E}_{loc,n}^{(i)} + \gamma^{(i)} \sum_{\substack{m=1 \\ m\neq n}}^{N} \boldsymbol{G}_{nm} \cdot \left(\boldsymbol{E}_{inc,m} + \boldsymbol{E}_{scatsum,m}^{(i)}\right) + \delta^{(i)} \sum_{\substack{m=1 \\ m\neq n}}^{N} \boldsymbol{G}_{nm} \cdot \boldsymbol{E}_{loc,m}^{(i)}$$

(9)

which allows for more consistent and better convergence. The optimization parameters $\alpha^{(i)}, \beta^{(i)}, \gamma^{(i)}$, and $\delta^{(i)}$ are calculated for each iteration step $i$ using the same principle as in the original CSDDA, i.e. by minimizing the upgraded relative error $R^{(i+1)}$,

$$R^{(i+1)} = \sqrt{\frac{\sum_{n=1}^{N}|\boldsymbol{E}_{inc,n} + \boldsymbol{E}_{scatsum,n}^{(i+1)} + \sum_{\substack{m=1 \\ m\neq n}}^{N}\boldsymbol{G}_{nm}\cdot(\boldsymbol{E}_{inc,m} + \boldsymbol{E}_{scatsum,m}^{(i+1)}) - \boldsymbol{E}_{loc,n}^{(i+1)} - \sum_{\substack{m=1 \\ m\neq n}}^{N}\boldsymbol{G}_{nm}\cdot\boldsymbol{E}_{loc,m}^{(i+1)}|^2}{N}}.$$

(10)

We thus obtain a linear algebraic equation in the 4 unknowns $\alpha^{(i)}, \beta^{(i)}, \gamma^{(i)}$, and $\delta^{(i)}$ for each iteration step $i$, which then updates the local electric field at each dipole $n$.

Finally, we also include the impact of the substrate on which the nanocrystals are grown. This can be approximated with good accuracy through the Green's function of an infinite plate (the NC is placed just one dipole thickness above the plate). Then for each dipole $n$, we have a scattering term $\boldsymbol{G}_{nn}$ that contributes to the iterative scheme so that $m = n$ contributions should also be included,

$$\boldsymbol{E}_{loc,n}^{(i+1)} = \alpha^{(i)}\left(\boldsymbol{E}_{inc,n} + \boldsymbol{E}_{scatsum,n}^{(i)}\right) + \beta^{(i)}\boldsymbol{E}_{loc,n}^{(i)} + \gamma^{(i)} \sum_{m=1}^{N} \boldsymbol{G}_{nm} \cdot \left(\boldsymbol{E}_{inc,m} + \boldsymbol{E}_{scatsum,m}^{(i)}\right) + \delta^{(i)} \sum_{m=1}^{N} \boldsymbol{G}_{nm} \cdot \boldsymbol{E}_{loc,m}^{(i)}.$$

(11)

Note that the term $m = n$ is also included in the summations. The Green's function has the following expression depending on values of $m$ and $n$ relative to each other[13–15],

$$\boldsymbol{G}_{nm} \cong \boldsymbol{G}(\boldsymbol{r}_n, \boldsymbol{r}_m, \omega) = \frac{e^{ik\rho}}{4\pi\varepsilon_0\rho^3}[-\{1 - k\rho - (k\rho)^2\}\boldsymbol{I} + \{3 - 3ik\rho - (k\rho)^2\}\frac{\rho\otimes\rho}{\rho^2}], m \neq n, \quad (12)$$

$$\boldsymbol{G}_{nn} \cong \boldsymbol{G}(\boldsymbol{r}_n, \boldsymbol{r}_n, \omega) = \frac{1}{32\pi\varepsilon_0 z_n^3}\frac{\varepsilon(\omega)-1}{\varepsilon(\omega)+1}\begin{pmatrix} 1 & 0 & 0 \\ 0 & 1 & 0 \\ 0 & 0 & 2 \end{pmatrix}, \quad m = n \quad (13)$$

Due to the strong absorption in most of the spectral domain, it is a reasonable approximation to consider the Green's functions in the near-field limit $(k\rho) \ll 1$. The final results do not differ from calculations for the full Green's function, but due to limited propagation effects, offer faster convergence at the same time. Also, in Gaussian units, $\varepsilon_0 = 1/4\pi$.

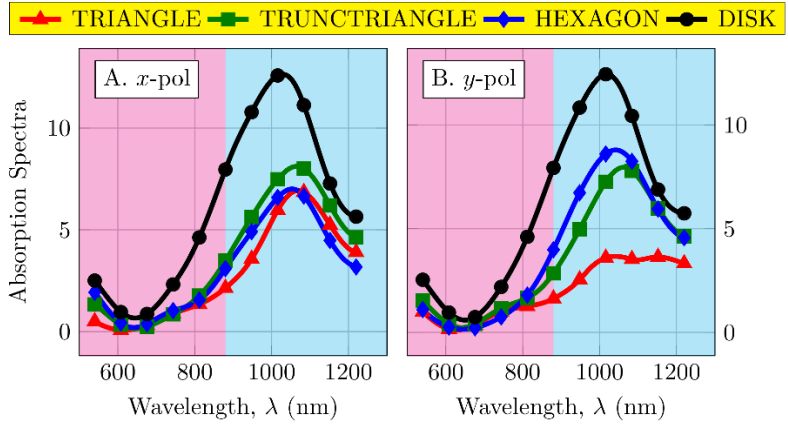

**Fig. S6** Intensity-normalized absorption spectra for CuSe NCs of different shapes, with same diagonal length $a$ = 19 nm, and thickness $c$ = 5 nm, for incident field polarized along the (A) x-axis and (B) y-axis, $E_{inc} = e^{ikz}$ (total number of dipoles, $N_d = 80 \times 80 \times 20 = 128{,}000$).

**CSDDA simulations for different angles of incidence**

By changing the angle of incidence from $\theta_k = 0$ to $\theta_k = 60$ degrees, we can gradually increase the LSPR absorption cross-section in the hyperbolic spectral range (Fig. S7 and Fig. S8). At $\theta_k = 75$ degrees, the hyperbolic LSPR dominates entirely over the metallic LSPR – we observe this effect across all crystal shapes. The disk NC (shown here for comparison purpose), due to its high symmetry, has nearly identical absorption spectra in Fig. 8 and 9, for a given incidence angle $\theta_k$ (the minor discrepancy for $\theta_k = 45$ degrees are likely due to DDA errors).

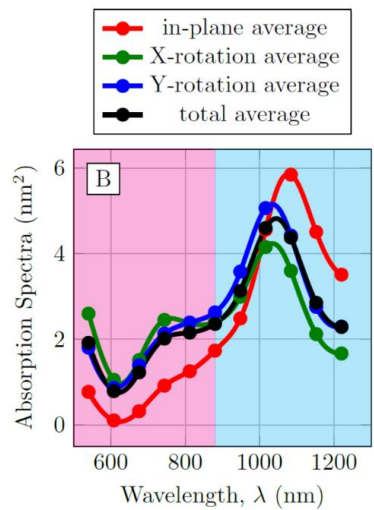

**Fig. S7** Rotational averaging of the absorption spectra for changing polarization in-plane of the NC surface, from CSDDA++. The hyperbolic domain is shown in pink.

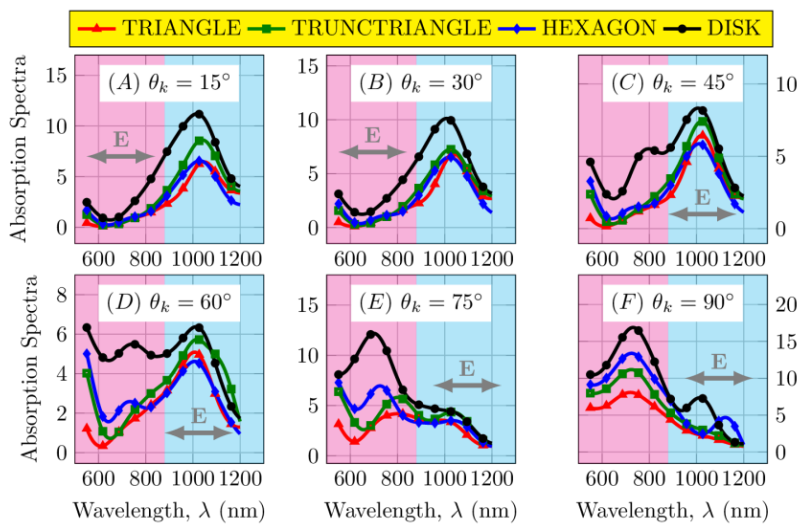

**Fig. S8** Intensity-normalized absorption cross-section for triangle (red triangles), truncated triangle (hexatrian) (green squares), hexagon (blue diamond), and disk (magenta circles) NC geometry, with total number of dipoles, $N_d = 80 \times 80 \times 20 = 128{,}000$. The incident field is polarized along the *x*-axis, with angle of incidence $\theta_k$ (A-F). Absorption cross-section can be enhanced to as high as 17 nm² (F), by controlling the angle of incidence. Gradually increasing the incidence angle also amplifies the LSPR in the hyperbolic regime, while suppressing the LSPR in the metallic domain. By $\theta_k = 90$ degrees (F), the plasmon resonances look qualitatively identical for every pr shape.

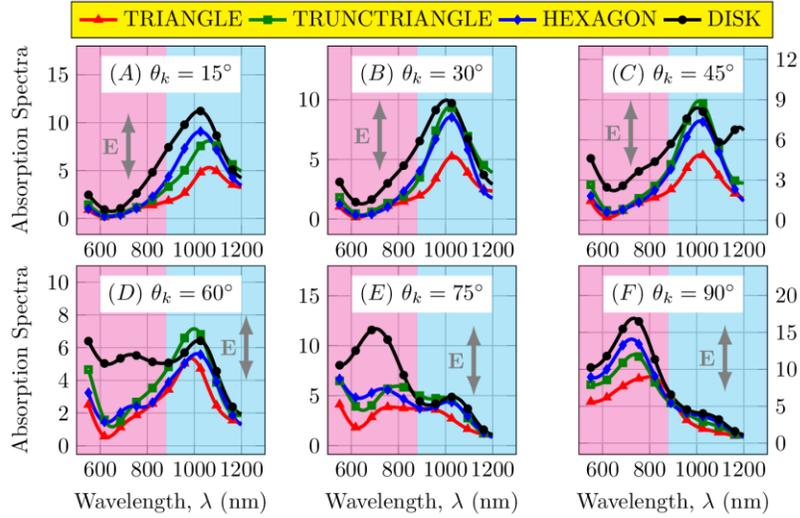

**Fig. S9** Intensity-normalized absorption cross-section for triangle (red triangles), truncated triangle (hexatrian) (green squares), hexagon (blue diamond), and disk (magenta circles) NC geometry, with total number of dipoles, $N_d = 80 \times 80 \times 20 = 128{,}000$. The incident field is polarized along the *y*-axis, with angle of incidence $\theta_k$ (A-F). Absorption cross-section can be enhanced to as high as 15 nm² (F), by controlling the angle of incidence. Gradually increasing the incidence angle also amplifies the LSPR in the hyperbolic regime, while suppressing the LSPR in the metallic domain By $\theta_k = 90$ degrees (F), the plasmons look almost identical for every crystal shape.

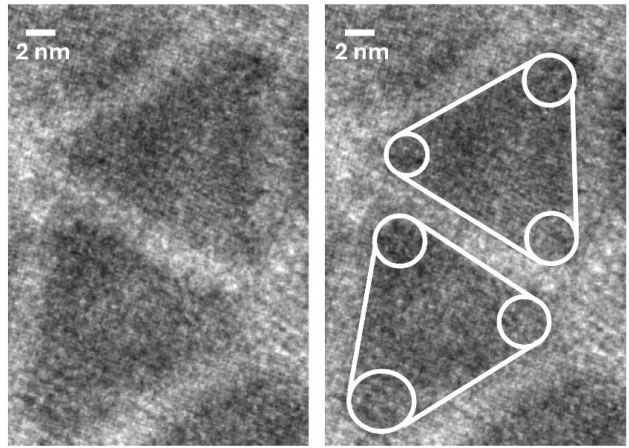

**Fig. S10** TEM image of triangular CuSe nanoprisms without (left) and with (right) alignment of the vertices with overlayed radial curvatures.

The impact of the intrinsic material anisotropy can be deciphered, by comparing our field distribution maps in Fig. 11, to the corresponding case of *isotropic* triangular nanoprism in Figs. 12a and 12b. First, we assume a triangular nanoprism with the same shape and dimensions as in Fig. S11, but with isotropic permittivity $\varepsilon_{xx}$, which remains metallic throughout the spectral range of interest. As seen in Fig. S12a, we obtain the standard metallic response with strong dipolar and quadrupolar LSPRs and high amplification. However, at no wavelength do we observe the "enveloping" mode that appears at $\lambda = 773$ nm in the anisotropic case. Similarly, we consider an isotropic triangular nanoprism with permittivity $\varepsilon_{zz}$ applied both in-plane and out-of-plane (Fig. S12b). Here, the optical response gradually shifts from dielectric to metallic. In the dielectric regime, the plasmonic excitations tend to "leak" into the environment, and even at $\lambda = 773$ nm the isotropic nanoprism remains weakly dielectric. As the wavelength enters the NIR range, the $\varepsilon_{zz}$-nanoprism becomes metallic, but only edge-confined quadrupolar excitations appear, with gradual extinction of the plasmonic response. In contrast, the anisotropic hyperbolic nanoprism (Fig. S11b) supports strongly confined excitations and clearly exhibits the full-surface "enveloping" mode. Neither isotropic case reproduces this effect or the considerably stronger enhancement seen in the anisotropic structure, which results from its in-plane metallic excitations. We thus see direct evidence that the enveloping modes are a direct consequence of intrinsic material anisotropy.

In conclusion, shape anisotropy is responsible for blue shifts, hotspots formation, asymmetric spectral broadening (as one moves from disk to hexagon to truncated triangle to triangle geometry in Figs. S8 and 9 for any incidence angle $\theta_k$), while the intrinsic material anisotropy is directly responsible for enveloping modes shown close to epsilon near zero (ENZ) wavelengths in a naturally hyperbolic material.

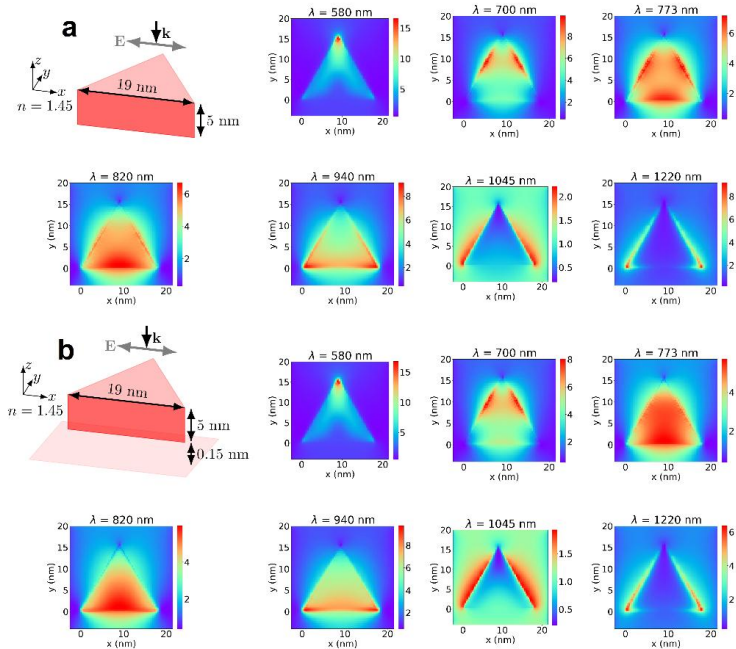

**Fig. S11** Substrate effect on the E-field distribution in triangular CuSe nanocrystal (CSDDA simulation). (a) NC in a solvent without the substrate (repeated for comparison). (b) NC in a solvent on a substrate.

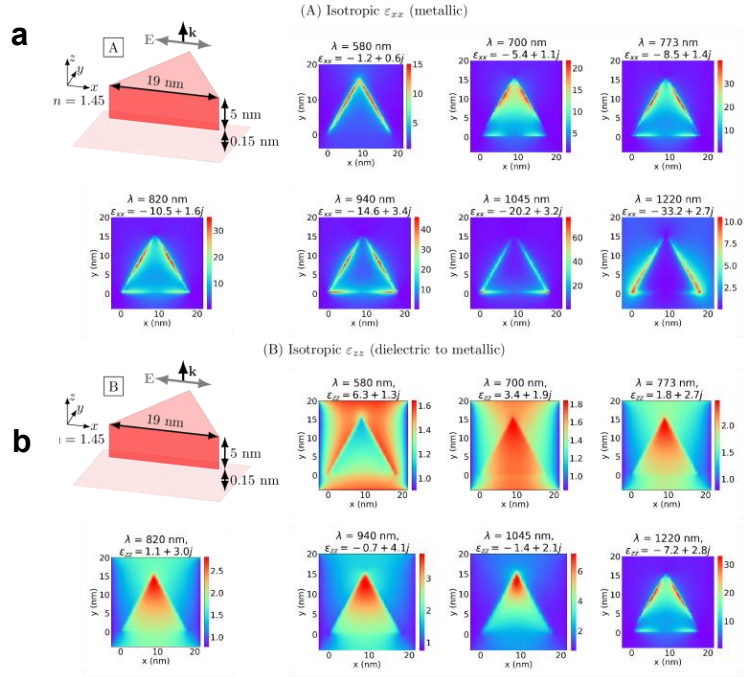

**Fig. S12** Intensity-field profile around an "model" triangular nanocrystal with isotropic permittivity for (a) $\varepsilon_{xx}$-isotropy and (b) $\varepsilon_{zz}$-isotropy. Comparing with Fig S11b, we can see the impact of the material anisotropy.

## 4. Fitting of TA oscillatory features

The averaged time traces were fitted using a model consisting of two exponential decay terms and an oscillatory component:

$$y = A_1 \cdot e^{\frac{-t}{\tau_1}} + A_2 \cdot e^{\frac{-t}{\tau_2}} + B \cdot e^{\frac{-t}{\tau_d}} \cdot \cos\bigl((2\pi f \cdot t) + \varphi \bigr)$$

where the amplitudes $A_1$, $A_2$ and the time constants $\tau_1$ und $\tau_2$ describe the exponential decay components, while B, $\tau_d$, f and $\varphi$ represent the amplitude, damping constant, frequency and phase of the oscillation, respectively. The relevant extracted values are summarized in Table 1 of the main text. The oscillation frequency obtained from the fit is 7.5 THz for NSs and 7.8 THz for NPLs, which is in good agreement with the FFT results.

## 5. Fourier Transform of TA data for different spectral regions

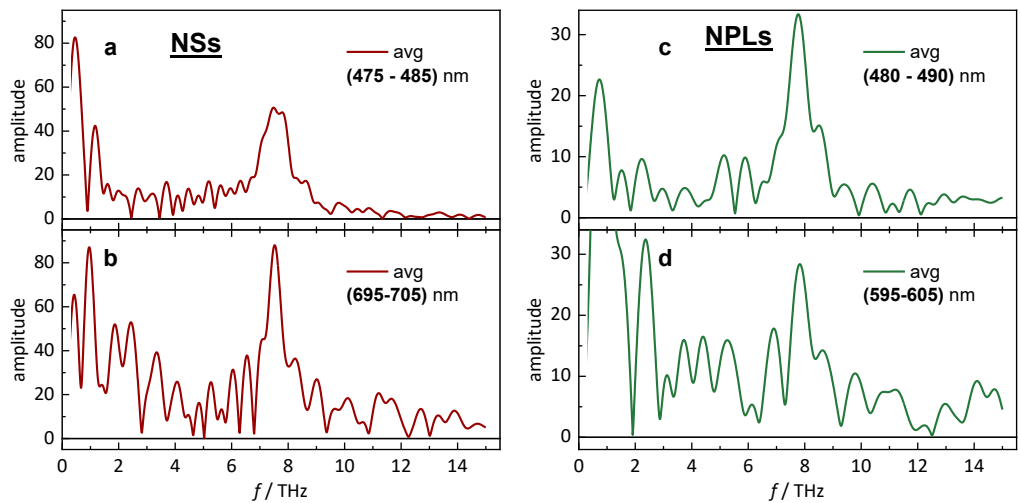

**Fig. S13 (a) and (c)** FFT data of the NSs and NPLs averaged over a spectral window in the flank of the UV absorption. **(b) and (d)** FFT data of the NSs and NPLs averaged over a spectral window at the absorption minimum of the stationary absorption spectra. For both types of particles, the Fourier amplitudes of the oscillations in the TA signal are similar in both spectral windows.